\documentclass[fleqn,usenatbib]{mnras}

\usepackage{newtxtext,newtxmath}

\usepackage[T1]{fontenc}

\DeclareRobustCommand{\VAN}[3]{#2}
\let\VANthebibliography\thebibliography
\def\thebibliography{\DeclareRobustCommand{\VAN}[3]{##3}\VANthebibliography}

\usepackage{tabularx}
\usepackage{subcaption}
\usepackage{lineno}

\usepackage{graphicx}	
\usepackage{amsmath}	

\newcommand{\msun}{${\rm M}_{\odot}$}

\DeclareRobustCommand*{\th}{$^{\textrm{th}}$}

\title[The atomic gas sequence and MZR of galaxies]{The atomic gas sequence and mass-metallicity relation from dwarfs to massive galaxies}

\author[D. Scholte et al.]{
Dirk Scholte,$^{1,2}$\thanks{E-mail: dscholte@ed.ac.uk (DS)}
Am\'elie Saintonge,$^{1}$
John Moustakas,$^{3}$
Barbara Catinella,$^{4,5}$
Hu Zou,$^{6}$
Biprateep Dey,$^{7}$
\newauthor{
J.~Aguilar,$^{8}$ 
S.~Ahlen,$^{9}$
A.~Anand,$^{8}$
R.~Blum,$^{10}$ 
D.~Brooks,$^{1}$ 
C.~Circosta,$^{1}$ 
T.~Claybaugh,$^{8}$ 
} 
\newauthor{
A.~de la Macorra,$^{11}$ 
P.~Doel,$^{1}$
A.~Font-Ribera,$^{1,22}$
P.~U.~F\"orster,$^{1}$ 
J.~E.~Forero-Romero,$^{12,13}$ 
} 
\newauthor{
E.~Gaztañaga,$^{14,15,16}$ 
S.~Gontcho A Gontcho,$^{8}$
S.~Juneau,$^{10}$
R.~Kehoe,$^{17}$
T.~Kisner,$^{8}$ 
S.~E.~Koposov,$^{2,31}$
}
\newauthor{
A.~Kremin,$^{8}$ 
A.~Lambert,$^{8}$ 
M.~Landriau,$^{8}$
C.~Maraston,$^{15}$
P.~Martini,$^{18,19,20}$ 
A.~Meisner,$^{10}$
A.~S.~Mighty,$^{1}$
} 
\newauthor{
R.~Miquel,$^{21,22}$ 
A.~D.~Myers,$^{23}$ 
J.~Nie,$^{6}$ 
C.~Poppett,$^{8,24,25}$  
F.~Prada,$^{32}$
M.~Rezaie,$^{26}$ 
G.~Rossi,$^{27}$
}
\newauthor{
E.~Sanchez,$^{28}$
M.~Schubnell,$^{29,30}$
J.~Silber,$^{8}$ 
D.~Sprayberry,$^{10}$
M.~Siudek,$^{16,22}$ 
F.~Speranza,$^{1}$ 
G.~Tarl\'{e} $^{30}$ 
}
\newauthor{
and B.~A.~Weaver $^{10}$
}
\\
\\
Author affiliations are listed at the end of the paper.
}

\date{Accepted XXX. Received YYY; in original form ZZZ}

\pubyear{\the\year{}}

\begin{document}
\label{firstpage}
\pagerange{\pageref{firstpage}--\pageref{lastpage}}
\maketitle

\begin{abstract}
Galaxy scaling relations provide insights into the processes that drive galaxy evolution. The extension of these scaling relations into the dwarf galaxy regime is of particular interest. This is because dwarf galaxies represent a crucial stage in galaxy evolution, and understanding them could also shed light on their role in reionising the early Universe. There is currently no consensus on the processes that dominate the evolution of dwarfs. In this work we constrain the atomic gas sequence (stellar mass vs. atomic gas fraction) and mass-metallicity relation (stellar mass vs. gas phase metallicity) from dwarf ($10^{6.5}$ \msun) to massive ($10^{11.5}$ \msun) galaxies in the local Universe. The combined optical and 21-cm spectroscopic observations of the DESI and ALFALFA surveys allow us to simultaneously constrain both scaling relations. We find a slope change of the atomic gas sequence at a stellar mass of $\sim 10^{9} ~\textrm{M}_{\odot}$. We also find that the shape and scatter of the atomic gas sequence and mass-metallicity relation are strongly linked for both dwarfs and more massive galaxies. Consequently, the low mass slope change of the atomic gas sequence is imprinted onto the mass-metallicity relation of dwarf galaxies. The mass scale of the measured slope change is consistent with a predicted escape velocity threshold below which low mass galaxies experience significant supernova-driven gas loss, as well as with a reduction in cold gas accretion onto more massive galaxies. 
\end{abstract}




\section{Introduction}

Galaxy scaling relations are the observable result of a complex network of processes driving galaxy evolution.  Many of the processes that dominate the evolution of galaxies relate to the accretion, processing and ejection of gas from galaxies \citep{tinsley1980}; combined, these processes are referred to as the baryon cycle. Many of the elements of the baryon cycle, in particular the gas inflow and outflow rates, are unfortunately challenging to accurately measure \citep{tumlinson2017}. This is a significant hurdle in the understanding of galaxy evolution through the baryon cycle, given the central importance of these processes. Luckily however, there is significant information about all the components of the baryon cycle encoded in the much more readily observable cold interstellar medium of galaxies \citep{saintonge2022}.

Indeed, the connection between various observables of galaxies through processes in the baryon cycle is reflected in the correlation of the atomic and molecular gas mass of galaxies with other galaxy properties, such as luminosity, stellar mass, baryonic mass, size, morphology and star formation rate \citep[e.g.][]{tully1977, kenney1988, sage1993, roberts1994, brinchmann2004, bothwell2013, bothwell2016, lelli2017, lelli2019, bacchini2020}. Large surveys of the atomic \citep[e.g. xGASS;][]{catinella2018} and molecular \citep[e.g. xCOLDGASS;][]{saintonge2017} cold gas content of representative samples of galaxies established gas scaling relations and quantified the role of gas in shaping the observed galaxy distributions as a function of their main observable properties. A clear connection between the gas within galaxies and the circumgalactic medium (CGM) has also been established; showing that the HI disk of galaxies is fueled by the CGM \citep{borthakur2015}; this further strengthens the picture of the interaction of inter- and circumgalactic gas with the gas phase of galaxies through the baryon cycle.

Baryon cycle processes are closely related to the chemical evolution of galaxies. Gas is the fuel of star formation and therefore an indirect driver for chemical enrichment. The accretion of (relatively) pristine gas onto galaxies also has a diluting effect on the gas phase metallicity of the interstellar medium (ISM) of galaxies. This dilution lowers the metallicity of galaxies that have accreted a large amount of gas, which is imprinted on the scatter of the mass-metallicity relation \citep[MZR; the relation between stellar mass and gas-phase metallicity;][]{tremonti2004}. In the nearby Universe, where gas mass measurements of representative samples of galaxies are available, the effect of metallicity dilution on the MZR has been measured \citep{brinchmann2013, bothwell2013, bothwell2016, lara-lopez2013, hughes2013, brown2018, scholte2023, marino2013}. These gas mass measurements are not readily available for large samples of galaxies at higher redshifts. Therefore the scatter (and redshift evolution) of the MZR has more often been studied as a function of star formation rate \citep[e.g.][]{ellison2008}. The three-parameter relation between stellar mass, gas phase metallicity and star formation rate is often referred to as the fundamental metallicity relation \citep[FMR;][]{mannucci2010, mannucci2011, brown2016, cresci2019, curti2020}. In this work we study the connection between the baryon cycle and the chemical evolution of galaxies using measurements of the atomic gas sequence and mass-metallicity relation. We refer to the atomic gas sequence as the relation between the stellar mass ($M_{\star}$) of galaxies and their atomic gas fraction ($f_{\textrm{HI}} \equiv M_{\textrm{HI}}/M_{\star}$).

Constraining galaxy scaling relations in the dwarf galaxy regime is of particular interest. Due to the many orders of magnitude lower mass of dwarfs compared to massive galaxies it is unclear whether feedback and gas accretion operate similarly over the entire stellar mass range covered in this work (from $10^{6.5}$ \msun{} to $10^{11.5}$ \msun{}). These questions so far remain mostly unanswered as observing large representative samples of dwarf galaxies is challenging due to their faintness. However, significant samples have been compiled \citep[e.g.][]{Karachentsev2019} in the Local Volume and nearby Universe, which suggest that he atomic gas fraction of dwarf galaxies is much higher than for massive galaxies \citep[e.g.][]{geha2006} and that their gas phase metallicities are systematically lower \citep{jimmy2015}. Understanding the physics of dwarf galaxies is also important due to the key role that dwarf galaxies may play in cosmic reionisation in the early Universe \citep[e.g.][]{atek2024, wu2024}.

We use observations from the Dark Energy Spectroscopic Instrument \citep{levi2013, desi2023} to study galaxy scaling relations far into the dwarf galaxy regime. DESI is a robotic, multiplexed spectroscopic instrument on the Mayall 4-meter telescope at Kitt Peak National Observatory \citep{desi2022}. DESI can obtain simultaneous spectra of almost 5000 objects \citep{desi2016b, silber2023, miller2023} and is currently conducting a five-year survey of about a third of the sky \citep{desi2016, schlafly2023}. The DESI survey is the largest spectroscopic galaxy survey to date; it will obtain spectra of $\sim$40 million unique galaxies and quasars at redshifts between 0.0 $<$ z $<$ 3.7 during a 5 year scientific programme \citep{desi2016, desi2024baogalaxies,desi2024baolya,desi2024baocosmo}. In particular, the faint magnitude limits of the DESI Bright Galaxy Survey (Bright: $r < 19.5$, Faint: $19.5 < r < 20.175$) and LOW-Z secondary target programme ($r < 21.0$) allow us to study a large sample of dwarf galaxies of the nearby Universe \citep{hahn2023, darragh-ford2023}. In this work we measure atomic gas masses using the large shared footprint between DESI and the Arecibo Legacy Fast ALFA Survey (ALFALFA), which is a blind extragalactic HI survey of the local universe \citep{giovanelli2005, haynes2018}. Using ALFALFA we directly measure the atomic hydrogen content of galaxies through measurements of the 21-cm hydrogen emission line. The combined DESI and ALFALFA data is a valuable homogeneous dataset of optical and 21-cm spectroscopy. Our measurements span 5 orders of magnitude in stellar mass from the dwarf galaxy regime ($10^{6.5}$ \msun) to massive galaxies ($10^{11.5}$ \msun), and we use them to perform a joint analysis of the atomic gas sequence and of the mass-metallicity relation.  

This paper is structured as follows: in Section \ref{sec:observations} we discuss the observations and data products used in this work. In Section \ref{sec:methods} we describe our analysis methods and in Section \ref{sec:results} we discuss the results. We discuss the implications and possible explanations for our results in Section \ref{sec:discussion} and summarise our main results in \ref{sec:conclusions}. We assume the cosmological parameters from the \cite{planck2020cmb}.

\section{Observations and data products}
\label{sec:observations}
Our main sample is a subset of the DESI Bright Galaxy Survey (BGS) validation and Year 1 Main Survey Data \citep{hahn2023bgs}. The Bright Galaxy Survey observes galaxies in the lowest redshift ranges of the DESI survey between 0.0 $<$ $z$ $<$ 0.6. The year 1 data set of the BGS includes extragalactic spectra for over 10 million galaxies. The spectrographs of the DESI instrument cover a wavelength range between 3600 \AA{} and 9800 \AA{}, split over three cameras. The resolving power, $R = \lambda/\Delta\lambda$, ranges from approximately 2000 at the shortest wavelengths to 5500 at the longest wavelengths \citep{desi2023}. The observed spectra are reduced using an extensive spectroscopic reduction pipeline \citep{guy2023}. Spectra are classified and redshifs are measured using a template fitting pipeline \citep[][Bailey et al. in prep.]{anand2024}. The DESI observations (including value added catalogs) used here will be made publicly available in DESI data release 1\footnote{\url{https://data.desi.lbl.gov/doc/}}. Some of the data is already publicly available as part of the DESI early data release \citep[EDR;][]{desi2023, desi2023edr}.

Our analysis makes use of several value added catalogs which have been produced for DESI. The photometric measurements used here are derived from the 9$^{\rm th}$ data release DESI Legacy Imaging Surveys \citep{dey2019}. The photometry is based on the Mayall z-band Legacy Survey (MzLS), Dark Energy Camera Legacy Survey (DECaLS) and Beijing-Arizona Sky Survey \citep[BASS;][]{zou2017}. The photometry consist of three optical bands (g, r, z), which are complemented by observations in four infrared bands (W1-4) taken from the Wide-field Infrared Survey Explorer \citep[WISE;][]{mainzer2014}.

The emission line fluxes in the DESI spectra are measured using \textsc{FastSpecFit}, a stellar continuum and emission line modelling code optimised for DESI, which jointly models the three-camera optical spectrophotometry from DESI and the legacy survey imaging \citep{moustakas2023}.

\begin{figure}
	\begin{center}
	\includegraphics[width=0.5\textwidth]{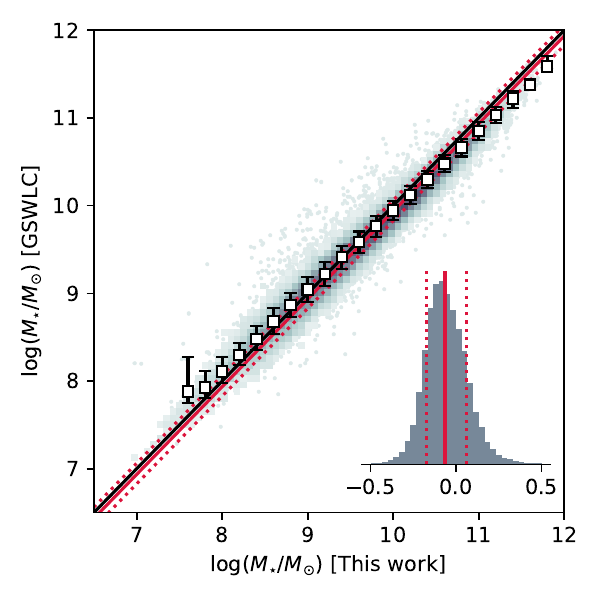}
	\caption{Comparison between the stellar masses used here and stellar masses from the GSWLC catalog \citep{salim2016,salim2018}. Our stellar mass measurements are on average $0.06^{+0.11}_{-0.13}$ dex higher than the GSWLC measurements. The black/white data points show the running median with errorbars showing the 16$\rm ^{th}$ and 84$\rm ^{th}$ percentile values. The red lines are not fits to the data but show the median and 16$\rm ^{th}$ and 84$\rm ^{th}$ percentile intervals over all the data. The black line shows the 1:1 relation. The histogram on the bottom right shows the distribution of the residuals.}
	\label{fig:stellar_mass_fastspec_sdss}
	\end{center}
\end{figure}

The stellar mass measurements are derived from the spectra, Legacy Surveys g, r, z band photometry and WISE band 1 and 2 photometry using the same approach as in \cite{zou2024}. We use the \textsc{Code Investigating GALaxy Emission} version 2022.1 \citep[CIGALE;][]{boquien2019} to model the spectral energy distributions (SEDs) of galaxies using the \cite{bruzual2003} simple stellar populations and a \cite{chabrier2003} initial mass function. We show a comparison between the CIGALE stellar masses and the measurements from the GSWLC-X catalog (version 2) of SDSS galaxies for the objects in the overlap between the two samples in Figure \ref{fig:stellar_mass_fastspec_sdss} \citep{salim2016, salim2018}.

Aside from the DESI spectroscopic observations we also make use of observations from the Arecibo Legacy Fast ALFA Survey (ALFALFA) which is a blind extragalactic HI survey of the local Universe \citep{giovanelli2005, haynes2018}. Due to the significant overlap of the footprints of ALFALFA and DESI we were able to retrieve ALFALFA spectra capturing the hydrogen 21-cm emission for 69,282 galaxies in our sample (see Section \ref{sec:sample_selection}). 

\section{Methods}
\label{sec:methods}
\subsection{Sample selection}
\label{sec:sample_selection}
Our sample of galaxies is taken from the DESI Survey Validation and Year 1 observations \citep{desi2023, desi2023b}. To ensure we can capture the hydrogen 21-cm line in the ALFALFA observations we impose a redshift selection of $0.001 < z < 0.06$. We also impose several quality selections to remove any objects from our sample that are not galaxy spectra or for which the photometry may not be reliable. We only use the primary spectra in the DESI redshift catalog. This ensures that we use the best spectrum available of each galaxy: \texttt{ZCAT\_PRIMARY == True}. We also require \texttt{SPECTYPE == GALAXY}, \texttt{ZWARN == 0} and \texttt{DELTACHI2 >= 40} to ensure we only include galactic spectra in our sample with succesful redshift measurements \citep{hahn2023bgs}. Using the photometric catalogs we limit the fraction of the flux in each of the optical bands coming from neighbouring objects, \texttt{FRACFLUX\_G/R/Z < 0.35}. This removes galaxies for which the observed flux is heavily contaminated by nearby sources. The remaining sample consists of 216,333 galaxies. We were able to retrieve ALFALFA observations for 69,282; the remaining DESI galaxies are outside of the ALFALFA survey footprint. In the rest of this section we will report both the numbers of the DESI sample and combined DESI and ALFALFA observations; the numbers of galaxies observed in both surveys are listed in [square brackets].

We visually inspect the Legacy Surveys imaging of every galaxy with $\log(M_{\star}/\textrm{M}_{\odot}) < 7.3$ and remove any spectra that are from star forming regions in large nearby galaxies misclassified as dwarf galaxies; these are called ``shreds''. Through this inspection we remove 125/[44] of the 2,459/[828] objects below this mass threshold. Above $\log(M_{\star}/\textrm{M}_{\odot}) = 7.3$ the number of shreds becomes a negligible fraction of the total sample. The remaining 216,208/[69,238] spectra form the base sample for our analysis.

\subsubsection{Necessity of using multiple sample selections}
The literature sample selection conventions for studies of the atomic gas sequence are different from studies of the mass-metallicity relation. In surveys of the atomic gas content of galaxies volume limited and mass complete samples are constructed which allow us to infer the average gas masses of galaxy populations \citep[e.g.][]{brown2015, catinella2018}. However, studies of the mass metallicity relation require sample selections based on the flux of several emission lines that are required for the metallicity measurement \citep[e.g.][]{tremonti2004, mannucci2010, yates2012,juneau2014, kashino2016, curti2020}. To allow comparison to literature studies as well as comparison between these two scaling relations we use multiple sample selections throughout this work. The two main subsamples in our analysis are (1) a mass complete sample and (2) a sample with detections of all the optical emission lines needed to derive the gas-phase metallicities for individual galaxies.

\begin{figure}
	\begin{center}
		\includegraphics[width=0.5\textwidth]{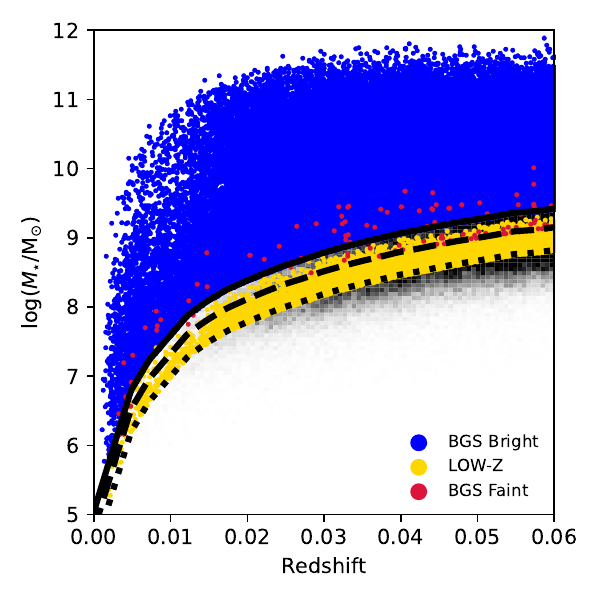} 
		\caption{The 90\% mass complete limits for the BGS Bright (blue data points, solid black line), Faint (red data points, dashed black line) and LOW-Z (yellow data points, dotted black line) sub-samples.}
		\label{fig:mass_complete_sample_selection}
	\end{center}
\end{figure}

\subsubsection{Mass complete sample} 
We determine a 90\% mass complete sample of galaxies following a similar procedure as in \cite{pozzetti2010}. We determine the mass each galaxy would have if its magnitude was equal to the magnitude limit of the survey: $\log(M_{\textrm{lim}}) = \log(M_{\star, i}) + 0.4 \times (m_{i} - m_{\textrm{lim}})$. Following this we select the faintest 20\% of galaxies in the sample, and in redshift bins of 0.025 dex we determine the mass below which 90\% of the limiting mass measurements of these faint galaxies fall. We define a continuous function $M_{\textrm{lim}}(z)$ by interpolating between the mass limits in each of the redshift bins. We choose the 90\% completeness limit following \cite{hahn2023}. 

Our DESI observations are divided into three categories with different magnitude limits. There is BGS Bright with $r < 19.5$ containing 185,959 galaxies, BGS Faint with $19.5 < r < 20.175$ (8,222 galaxies) and the LOW-Z secondary target programme with $r < 21.0$ (22,027 galaxies) \citep{hahn2023bgs,darragh-ford2023}. We compute the mass completeness limits for each of these categories. Our mass complete sample includes the 121,918/[39,869] galaxies above the mass completeness limit of the respective categories. The mass completeness selection on these different DESI observations are shown in the redshift -- stellar mass plane in Figure \ref{fig:mass_complete_sample_selection}.

\subsubsection{Emission line flux limited sample} 
To measure reliable gas phase metallicities we require a S/N $>$ 3 for the flux in the [\textsc{Oii}]$\lambda\lambda$3726, 3729, H$\beta$, [\textsc{Oiii}]$\lambda\lambda$4959, 5007, H$\alpha$, [\textsc{Nii}]$\lambda\lambda$6548, 6584 emission lines. We also require the line excitation to be caused by star formation. To ensure this we only include galaxies classed as star forming using the \cite{kauffmann2003} criterion:
\begin{equation}
    \log\left( [\textrm{\textsc{Oiii}}]/\textrm{H}\beta \right) \leq \frac{0.61}{\log\left( [\textrm{\textsc{Nii}}]/\textrm{H}\alpha \right) -0.05} + 1.3
\end{equation}
The remaining sample after these selections includes 76,726/[24,508] galaxies.

\subsection{Stacking procedure of DESI spectra}
We stack spectra of galaxies to derive high signal-to-noise emission line measurements for subsamples of our data set. This allows us to mitigate selection biases against galaxies with weak line emission. Our stacking procedure consists of the following steps:
\begin{enumerate}
	\item The spectra are corrected for Milky Way dust attenuation using galactic extinction measurements from \cite{schlegel1998} using a \cite{fitzpatrick1999} extinction curve with R$_V$ = 3.1. 
	\item Each spectrum is resampled to a common rest wavelength grid 
	(3600\AA -- 9800\AA, $\Delta\lambda$ = 0.2\AA) using the measured spectroscopic redshifts. We resample the spectra using linear interpolation \citep[e.g.][]{andrews2013}.
	\item Each spectrum is normalised using the flux of the H-alpha emission line.
	\item The spectra are then co-added, where the co-added flux is given by the mean flux in each wavelength bin. We co-add the flux in 200 bootstrap samples where the final co-added flux values are the mean flux measurements over the bootstrap samples of each wavelength bin.
	\item The co-added inverse variance is derived from the propagated inverse variance measured for each wavelength bin of each spectrum. The co-added inverse variance values we report are the mean inverse variances of the 200 bootstrap samples of each wavelength bin. 
\end{enumerate}
We experimented with several different normalisation (e.g. by luminosity or continuum flux) and weighting schemes (e.g. inverse variance or root mean square weighting) for the spectral stacking. On test samples with known metallicities of galaxies entering the stacks the current stacking algorithm most accurately produced stacks with measured metallicities representative of the sample. We measure the emission lines in the co-added spectra using the \texttt{stackfit} routine of \texttt{fastspecfit-v2.4}. This applies the same emission line fitting procedure which was used to derive the emission line measurements for the FastSpecFit value added catalog\footnote{\url{https://fastspecfit.readthedocs.io/en/latest/index.html}}.

\subsection{Metallicity measurement}
\label{sec:ch4_metallicity_measurement}
We derive gas-phase metallicities for the galaxies in our sample using the R23 (\{[O\textsc{ii}]3726,3729 + [O\textsc{iii}]4959,5007\}/H$\beta$) and N2 ([N\textsc{ii}]6584/H$\alpha$) strong line diagnostics \citep{pagel1979, storchi-bergmann1994, calzetti1994}. The measured R23 and N2 values are corrected for ISM dust attenuation using the \cite{cardelli1989} attenuation prescription and an assumed intrinsic H$\alpha$/H$\beta$ flux ratio of 2.86 based on case-B recombination \citep{osterbrock2006}. We use the calibration from \cite{nakajima2022} for the R23 diagnostic:
\begin{equation}
    \textrm{R23} = 0.515 - 1.474 x - 1.392 x ^2 - 0.274 x^3 ,
\end{equation}
where $x = 12 + \log(\textrm{O/H}) - 8.69$. To break the degeneracy in the R23 diagnostic we use the N2 diagnostic to derive unique solutions for the metallicities. We use the calibration from \cite{denicolo2002}, which can be written as:
\begin{equation}
	\textrm{N2} = - 0.589 + 1.370 x .
\end{equation}
The calibration by \cite{denicolo2002} is described by a linear relation between the N2 diagnostic and metallicity which is in agreement with the expected theoretical relation for this diagnostic over the metallicity range covered as well as with directly measured metallicities. We derive the metallicity using a least squares approach where we jointly minimize the residuals for the R23 and N2 diagnostics. 

\begin{figure}
    \begin{center}
    \includegraphics[width=0.5\textwidth]{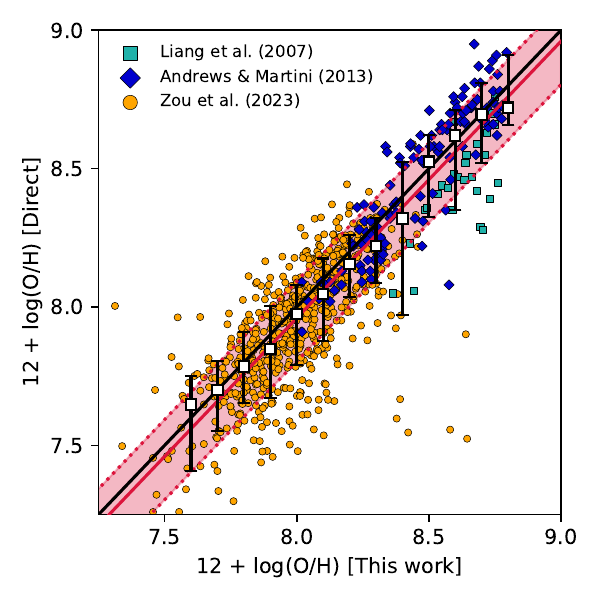} 
    \caption{A comparison between direct metallicity measurements and our strong line metallicity measurements. The direct metallicity measurements by \protect\cite{liang2007}, \protect\cite{andrews2013} and \protect\cite{zou2024} are shown in cyan squares, dark blue diamonds and orange circles, respectively. The black/white data points show the running median with errorbars showing the 16\th{} and 84\th{} percentile values. The red lines are not fits to the data but show the median and 16\th{} and 84\th{} percentile intervals over all the data. The black line shows the 1:1 relation.}
    \label{fig:metallicity_calibration}
    \end{center}
\end{figure}

We validate our metallicity measurements by comparing them to metallicities derived using measurements of the [O\textsc{iii}]4363 and/or [O\textsc{ii}]7320,30 auroral emission lines, which constrain the electron temperature and elemental abundances. We do this comparison using three samples with such temperature-sensitive metallicity measurements: the stacked SDSS spectra derived by \cite{liang2007} and \cite{andrews2013} for the higher end of the metallicity range, and the individual galaxies in the DESI Early Data Release from \cite{zou2024} cover the low metallicity range.   We expect one-to-one agreement between our chosen strong line metallicities and these direct measurements because the strong line diagnostics were calibrated using similar metallicity measurements. Figure \ref{fig:metallicity_calibration} shows that our strong line metallicities are in agreement with the direct metallicity measurements with a scatter of $\sim 0.3$ dex. Due to the small aperture of the fibers of the DESI instrument (diameter $\sim 1.5$ arcsec; approx. 0.3 -- 1.8 kpc in our redshift range) the observed spectra only capture the emission from the central regions of the galaxies. Therefore, the derived metallicities are representative of these central regions. This may introduce some biases towards higher metallicity in central regions through galactic metallicity gradients, however, this should not affect the overall results as discussed by e.g. \cite{andrews2013} and \cite{brown2018}.  

\subsection{Stacking procedure of ALFALFA spectra and $\rm M_{HI}$ measurement}
\label{sec:methods_alfalfa}

\begin{figure}
    \begin{center}
    \includegraphics[width=0.5\textwidth]{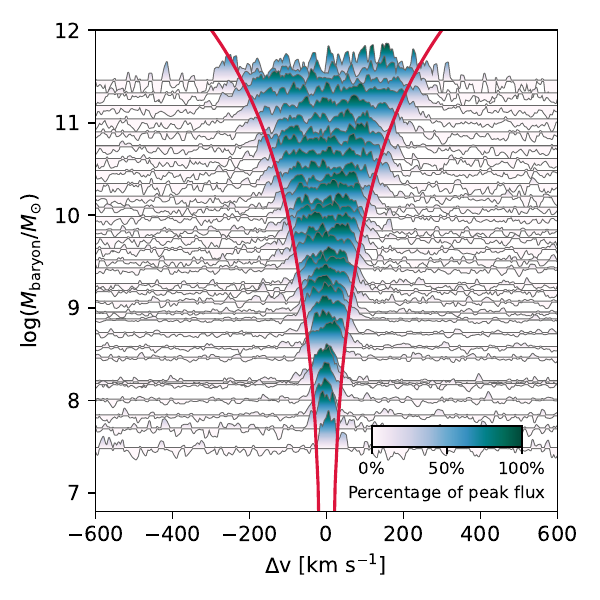} 
    \caption{The 21-cm line profiles of stacked spectra from our mass complete sample in stellar mass bins of 0.15 dex. On the x-axis we show the velocity offset from the 21-cm line and on the y-axis the zero flux point is centered on the median baryonic mass of each stack. In this figure we computed the baryonic mass as $M_{\textrm{baryon}} = M_{\star} + 1.33 \times M_{\textrm{HI}}$. The solid red line shows the baryonic Tully-Fisher relation measured by \protect\cite{lelli2019}, adjusted to our measurements which are not inclination corrected and including an added velocity dispersion due to the average DESI BGS redshift uncertainty \protect\citep{lan2023}.}
    \label{fig:stacked_alfalfa_spectra}
    \end{center}
\end{figure}

We measure the HI 21-cm line flux of galaxies in our sample using the ALFALFA survey \citep{giovanelli2005, haynes2018}. ALFALFA was conducted as a blind survey, which means that we can extract an ALFALFA spectrum from any position within the survey footprint. The survey covers the frequency range from 1335 to 1435 MHz, which means the HI 21-cm ($\sim$1420.41 MHz) line can be recovered at $z<0.06$. From the unresolved HI line profiles we can calculate the atomic gas mass of a galaxy:
\begin{equation}
\label{eq:mhi_calculation}
	M_{\textrm{HI}}[\textrm{M}_{\odot}] = \frac{2.356 \times 10^{5}}{(1+z)^{2}} [D_{L}(z)]^{2} \int S_{\nu} \textrm{d}\nu
\end{equation}
where $D_{L}(z)$ is the luminosity distance (in Mpc) to the galaxy at redshift, $z$, as measured from the HI spectrum in the observed velocity frame, and the integral is the HI-line flux density integrated over the line, with $S_{\nu}$ in units of Jy and $\nu$ in km s$^{-1}$ \citep{haynes1984}. These atomic gas masses are sensitive to the total gas mass of galaxies due to the large beam (diameter $\sim 3.5$ arcmin; approx. 50 -- 250 kpc) of the Arecibo telescope at these frequencies.

Most of our DESI-selected galaxies are un-detected in the HI line; this is highlighted by the fact that only $\sim$10\% of our sources are catalogued in \cite{haynes2018}. Therefore, we do not derive the $M_{\textrm{HI}}$ for individual galaxies but rather stack the ALFALFA spectra of several galaxies together to increase the signal-to-noise ratio of the measurement. Our stacking procedure for the ALFALFA observations has the following steps:
\begin{enumerate}
	\item The 21-cm emission line is centered for each of the spectra using the DESI redshift measurements and each of the ALFALFA spectra is resampled to a common rest-frequency grid with velocity offsets from --800 to 800 km s$^{-1}$ in intervals of 5 km s$^{-1}$. This is done for both the linear polarizations in the ALFALFA data.
	\item The flux arrays are converted to units of luminosity using the luminosity distance: $L_{\nu} = [D_{L}(z)]^{2} S_{\nu}^{\textrm{rest}}.$
	\item We perform median stacking, i.e. we take the weighted median of the luminosity in each velocity channel. The stacking weights are defined as: $w_{\textrm{stack}} = w/\sigma^2$, where $w$ are the ALFALFA spectral weights and $\sigma$ is the rms noise in the flux measurement.  This is done separately for each of the two linear polarisations. 
	 \item We combine the signal of the two linear polarizations by taking the mean flux in each  frequency bin.
	\item We integrate over the 21-cm emission line to measure the luminosity of the emission line. We determine the bounds to integrate over (width of the emission line) using the curve of growth.
	\item We convert the measured line luminosities to atomic hydrogen masses (see Equation \ref{eq:mhi_calculation} for the conversion).
	\item We derive measurement uncertainties using bootstrap resampling of the spectra. We resample the spectra with replacement 200 times and calculate the uncertainty on the derived atomic gas mass as the standard deviation of masses derived in the bootstrap samples. 
\end{enumerate}

We use a weighted median to derive our stacked measurements to reduce the impact from abnormalities in some of the spectra in the stacks. These outliers can, for example, be caused by source confusion where the atomic hydrogen content of a galaxy nearby the target contributes to the measured flux in the 21-cm line. 

We also measured the width at 20\% of the peak height of the 21-cm line ($W_{20}$). We correct the $W_{20}$ measurement for the line broadening due to the velocity dispersion induced by redshift uncertainty of DESI BGS \citep[$\sigma_{\textrm{BGS}}\sim$10 km s$^{-1}$,][]{lan2023} using the \citet{tully1985} correction method:
\begin{equation}
\begin{split}
    W_{\textrm{20}}^{2} = & ~ W_{\textrm{uncorr.}}^{2} + W_{\textrm{rand}}^{2} - 2W_{\textrm{uncorr.}}W_{\textrm{rand}}\left[ 1 - e^{-\left(W_{\textrm{uncorr.}}/W_{c}\right)^{2}} \right] \\ &
    - 2W_{\textrm{rand}}^{2} e^{-\left(W_{\textrm{uncorr.}}/W_{c}\right)^{2}}
\end{split}
\end{equation}
where $W_{\textrm{20}}$ is the corrected line width (at 20\% of the peak flux) due to the rotational velocity of the atomic gas. $W_{\textrm{uncorr.}}$ is the uncorrected line width directly measured from the stacks. $W_{\textrm{rand}}$ is the line width due to the random contribution to the line width; we assume $W_{\textrm{rand}} = 3.78 \times \sigma_{\textrm{BGS}}$ where the multiplication factor translates the 1-$\sigma$ width to width at 20\% of the peak flux based on a Gaussian profile. $W_{c}$ is the width where the 21-cm line profile changes from a Gaussian shape to a double-horned profile; we assume $W_{c}=120$. We do not perform an inclination correction before stacking the spectra, therefore we apply also an average inclination correction factor of $\sin(i) = 0.7$ in $W_{20}/(\sin(i))$, assuming a random distribution of inclinations of galaxies in the stacks.
These corrections allow us to compare our stacked results with line width measurements of individual galaxies. We use the width at 20\% of the peak flux as this reliably traces $V_{\textrm{max}}$ down to the dwarf galaxy regime \citep{sardone2023}, where $V_{\textrm{max}} \simeq W_{20,\textrm{corr}}/2$.

In Figure \ref{fig:stacked_alfalfa_spectra} we show the stacked spectra we produce for galaxies in 0.15 dex stellar mass bins. We overplot the expected velocity widths according to the measured baryonic Tully-Fisher relation \citep{tully1977, lelli2019}, where we have applied the corrections discussed above in reverse to provide a fair comparison between the stacked measurements and literature. On the y-axis we show the measured baryonic mass of the galaxies in each stack. This is calculated using  $M_{\textrm{B}} = M_{\star} + 1.33 \times M_{\textrm{HI}}$, where we multiply the atomic gas mass measurement by a factor of 1.33 to account for the contribution of Helium to the total gas mass \citep[e.g.][]{lelli2019}. The widths of our 21-cm line measurements align well with the maximum rotational velocities expected from the measured baryonic Tully-Fisher relation from \cite{lelli2019}.

\begin{table*}
	\caption{Parameter values for the fitted relations. These parameter values were determined using a weighted least squares minimization using the \textsc{emcee} MCMC algorithm \citep{foreman-mackey2013}.}
	\label{tab:fit_params}
\centering
\begin{tabular}{lllllll}
 \multicolumn{7}{c}{\textbf{Atomic gas sequence (Equation \ref{eq:atomic_gas_sequence})}}  \\ [0.05cm] \hline \hline
 Data sample & No. of galaxies & $\rm log\left(f_{\textrm{HI,7}}\right)$ & $\gamma_0$  & $\gamma_1$  & $\beta$ & $\rm log(M_0)$  \\ [0.05cm] \hline
 r-band flux limited & 69,238 & $0.85_{-0.04}^{+0.04}$ & $-0.06_{-0.07}^{+0.04}$ &  $-0.811_{-0.018}^{+0.016}$   & $1.10_{-0.21}^{+0.34}$  & $8.17_{-0.07}^{+0.10}$   \\ [0.05cm]
Line flux limited & 24,508 & $0.92_{-0.05}^{+0.05}$ & $-0.34_{-0.05}^{+0.08}$ &  $-0.88_{-0.08}^{+0.09}$   & $1.1_{-0.4}^{+1.0}$  & $9.50_{-0.19}^{+0.17}$  \\ [0.05cm] 
 Mass complete & 39,869 & $0.45_{-0.03}^{+0.04}$ & $-0.14_{-0.06}^{+0.07}$ &  $-0.854_{-0.033}^{+0.025}$   & $1.2_{-0.3}^{+0.4}$  & $8.98_{-0.11}^{+0.11}$  \\ [0.05cm] \hline 
 \multicolumn{7}{c}{\textbf{Mass-metallicity relation (Equation \ref{eq:mzr_curti})}} \\ [0.05cm] \hline \hline
 Data sample & No. of galaxies & $Z_0$ & $\beta$  & log(M$_{0}$)& $\gamma$ &   \\ [0.05cm] \hline
Line flux limited &  76,726 & $8.854_{-0.008}^{+0.009}$ & $1.20_{-0.09}^{+0.10}$ &  $10.49_{-0.03}^{+0.03}$   & $0.2439_{-0.0013}^{+0.0014}$  &  \\ [0.05cm] \hline
\end{tabular}
\end{table*}

\section{Results}
\label{sec:results}

\subsection{The atomic gas sequence}

\begin{figure*}
    \begin{center}
    \begin{subfigure}[b]{0.495\textwidth}
        \includegraphics[width=\textwidth]{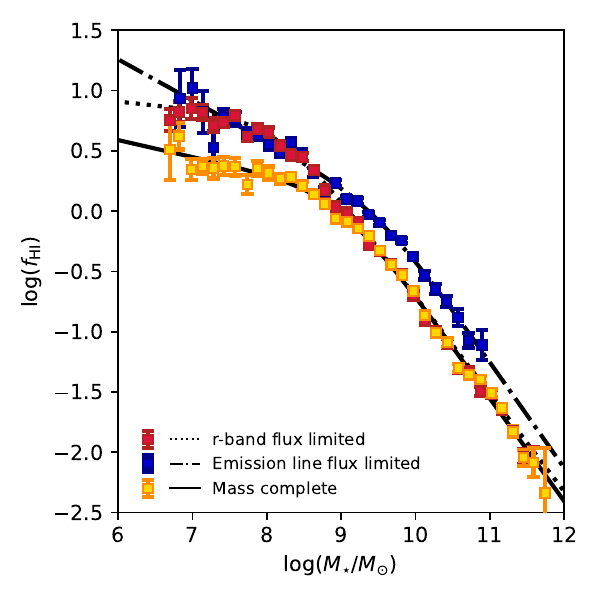}
    \end{subfigure}
    \hfill
    \begin{subfigure}[b]{0.495\textwidth}
        \includegraphics[width=\textwidth]{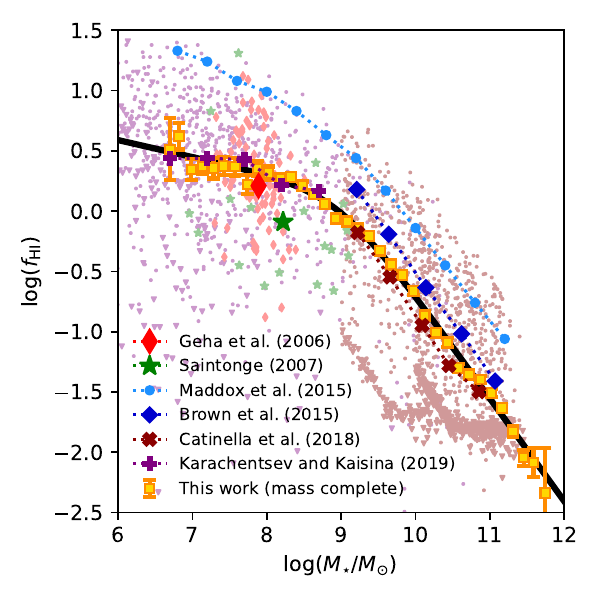}
    \end{subfigure}
    \caption{\textit{Left}: The atomic gas sequence as derived from stacked ALFALFA  spectra for the r-band flux limited (red, black dotted line), emission line flux limited (blue, black dash-dotted line) and mass complete samples (yellow, black solid line) together with the best fit relations from Equation \ref{eq:atomic_gas_sequence} as shown in Table \ref{tab:fit_params}. \textit{Right}: Literature comparisons to the sample from \protect\cite{geha2006} (red), the HI-selected sample from \protect\cite{saintonge2007} (green),  the ALFALFA HI-detections from \protect\cite{maddox2015} (light blue), stacked ALFALFA measurements \protect\citep[dark blue;][]{brown2015}, the xGASS representative survey of atomic hydrogen in galaxies \protect\citep[maroon;][]{catinella2018}, and the local volume sample from \protect\cite{Karachentsev2019} (purple). The large data points connected with dotted lines show the median atomic gas fractions within each survey, individual measurements are shown in the small data points and upper limits in small triangles.}
    \label{fig:fhi_vs_mstar}
    \end{center}
\end{figure*}

We measured the atomic gas fraction ($f_{\textrm{HI}} = M_{\textrm{HI}}/M_{\star}$) of the galaxies in our full ($r$-band flux limited), mass complete and emission line flux limited samples using stacked ALFALFA spectra in stellar mass bins of 0.15 dex. These measurements are shown in Figure \ref{fig:fhi_vs_mstar}. This figure shows a significant slope change for the atomic gas sequence between the dwarf and massive galaxy regimes. We parametrise the shape of the atomic gas sequence with a broken power law with a smooth transition region using the following functional form:
\begin{equation}
	\begin{split}
	\label{eq:atomic_gas_sequence}
	\log\left(f_{\textrm{HI}}\right) = & \log\left(f_{\textrm{HI,7}}\right)  + \gamma_0 \log\left(\frac{M_{\star}}{10^7}\right) \\ & + \frac{\gamma_1  - \gamma_0}{\beta} \log\left(1 + \left(\frac{M_{\star}}{M_{0}}\right)^{\beta}\right) .
    \end{split}
\end{equation}
In this equation $\log(f_{\textrm{HI,7}})$ is the atomic gas fraction at $M_{\star} = 10^7 ~\textrm{M}_{\odot}$, $\gamma_0$ is the slope at the low mass end, $\gamma_1$ is the slope at the high mass end, $\beta$ determines the width of the transition region between the two slopes and $M_0$ is the location of the transition. We fit the function to the data using a weighted least squares minimization. The best fit parameters were found using the \textsc{emcee} MCMC algorithm \citep{foreman-mackey2013}.

\begin{figure*}
    \begin{center}
    \begin{subfigure}{0.495\textwidth}
        \includegraphics[width=\textwidth]{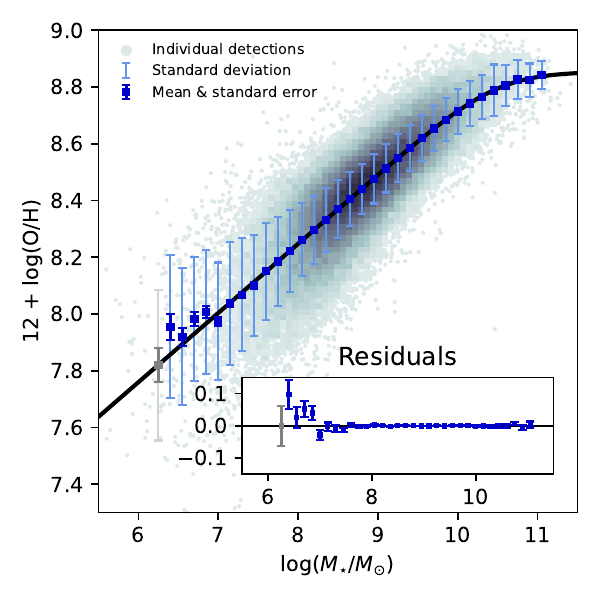}
    \end{subfigure}
    \hfill
    \begin{subfigure}{0.495\textwidth}
        \includegraphics[width=\textwidth]{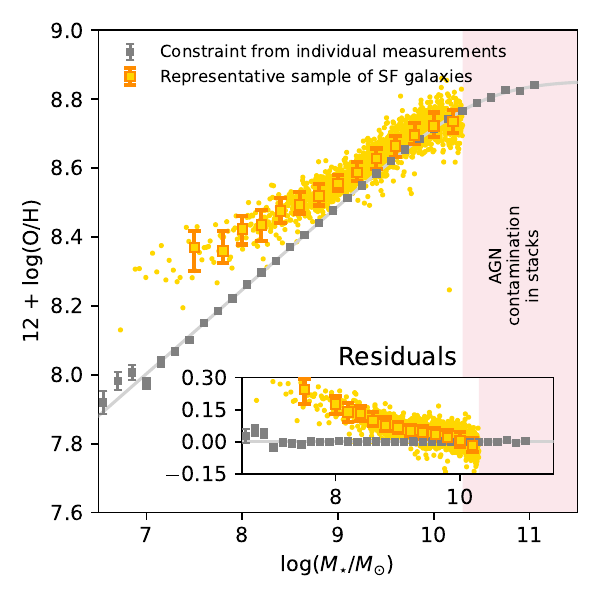}
    \end{subfigure}
    \caption{The mass-metallicity relation of star forming galaxies at $z<0.06$. \textit{Left}: This figure shows the emission line flux limited sample. The grey shaded area and datapoints show the individual measurements. The square data points show the mean mass-metallicity relation, standard error (dark blue/dark grey) and standard deviation (light blue/light grey) in 0.15 dex stellar mass bins for bins with 25 (blue)/10 (grey) or more galaxies. The fitted relation using Equation \ref{eq:mzr_curti} is shown in solid black. \textit{Right}: This figure shows the MZR for the mass complete stacks. The stacked measurements (yellow) are shown together with the fitted relation for the emission line flux limited sample (black) and binned averages of the stacked measurements (grey).}
    \label{fig:mzr}
    \end{center}
\end{figure*}

We compare our results of the atomic gas sequence to measurements from previous surveys in Figure \ref{fig:fhi_vs_mstar} (right panel). We compare to results from the $r$-band flux limited sample from \cite{geha2006}, the HI-selected sample from \cite{saintonge2007}, the ALFALFA sample of HI-detected galaxies from \cite{maddox2015}, stacked ALFALFA measurements from \cite{brown2015}, the xGASS representative galaxy survey of atomic gas masses \citep{catinella2018}, and the Local Volume sample composed by \cite{Karachentsev2019}. For the Local Volume sample we calculate the stellar masses from the reported K-band luminosity using a mass-to-light ratio of 0.21, which is the median mass-to-light ratio for our stellar mass measurements at $\log(M_{\star}) \sim 7.5$ in the neighbouring WISE-1 band. The large data points connected with dotted lines show the median measured atomic gas fraction as a function of stellar mass. 

The median results from these comparison samples are in good agreement with our measurements over the entire mass range. The sample of \cite{maddox2015} covers a wide stellar mass range, similar to our DESI sample, and detects a similar slope change at a stellar mass of $\sim10^9$ \msun.  As their sample only includes ALFALFA HI-detected galaxies, it is significantly biased towards the most gas-rich galaxies, explaining the significant offset, especially with our mass-complete sample. At stellar masses above $\sim10^9$\msun, we get excellent agreement with the median gas masses derived from the representative sample of the xGASS survey, as well as with the stacked ALFALFA measurements from \cite{brown2015}. In the dwarf galaxy regime we compare to samples from \cite{geha2006}, \cite{saintonge2007} and \cite{Karachentsev2019}. \cite{saintonge2007} derive slightly lower gas fractions compared to our measurements, however, here we have to take into account that their stellar mass calibrations are not matched to ours, which may introduce an artificial offset in both the x- and y-axis. The observed gas fractions and slope of the atomic gas sequence in the dwarf galaxy regime in the \cite{Karachentsev2019} sample is very similar to our measurements over the same mass range. We also note that a turnover of the atomic gas sequence around $\sim 10^9 ~\textrm{M}_{\odot}$ has previously been described \citep[e.g.][]{maddox2015, bradford2015}. 

\subsection{The Mass-metallicity relation}
We use both our emission line flux limited and our mass complete sample to study the mass-metallicity relation (MZR). The emission line flux limited sample is vital to study the scatter in the relation. However, the requirements on emission line signal-to-noise ratios introduce biases in the sample selection \citep{yates2012,juneau2014, kashino2016}. Therefore, we also use our mass complete sample to verify our measurement of the MZR. 

\subsubsection{The mass-metallicity relation for individual galaxies}
In the left panel of Figure \ref{fig:mzr} we show the MZR for the galaxies where we are able to derive gas-phase metallicities. We measure the mean metallicity and standard deviation in narrow 0.15 dex stellar mass bins, which is shown by the blue data points and error bars. The grey data point shows where we have only $<25$ galaxies in a bin. Our measurements constrain the MZR down to a stellar mass of $\sim 10^6 ~\textrm{M}_{\odot}$. We fit the functional form for the MZR from \cite{curti2020} to the binned stellar mass and metallicity measurements. This function is given by:
\begin{equation}
	\label{eq:mzr_curti}
	12 + \log(\textrm{O/H}) = Z_0 - \frac{\gamma}{\beta } \log\left(1 + \left(\frac{M_{\star}}{M_{0}} \right)^{-\beta } \right) .
\end{equation}
In this equation, $M_{\star}$ is the stellar mass, $Z_0$ the metallicity at the high mass end, $\gamma$ the power law index of the slope at low stellar masses, $\beta$ constrains the width of the transition region, and $M_{0}$ the turn-over stellar mass. We fit the function to the data using a weighted least squares minimization. The best fit parameters are shown in Table \ref{tab:fit_params} and were found using the \textsc{emcee} MCMC algorithm \citep{foreman-mackey2013}.

\begin{figure}
    \begin{center}
        \includegraphics[width=0.5\textwidth]{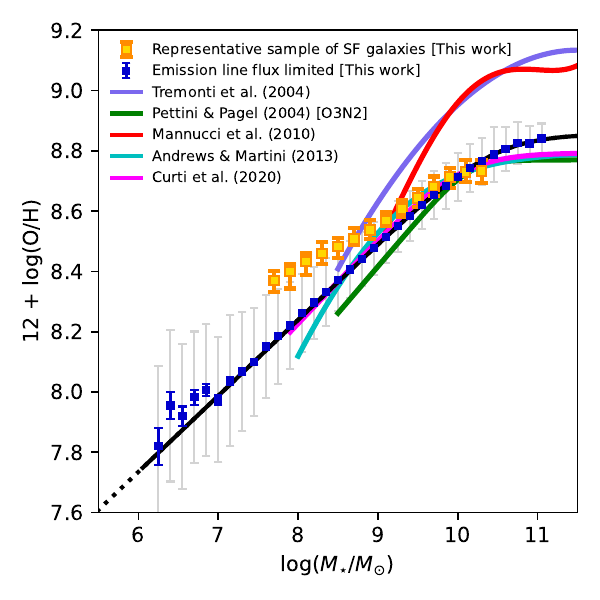} 
        \caption{Comparison of our MZR measurements (blue and orange; same as in Fig. \ref{fig:mzr}) to previous results by by \protect\cite{tremonti2004}, \protect\cite{pettini2004}, \protect\cite{mannucci2010}, \protect\cite{andrews2013} and \protect\cite{curti2020}. We find excellent agreement with the measurements from \protect\cite{andrews2013} and \protect\cite{curti2020}.}
        \label{fig:mzr_literature}
    \end{center}
\end{figure}

In Figure \ref{fig:mzr_literature} we compare our results to other measurements of the MZR at low redshift by \cite{tremonti2004}, \cite{pettini2004}, \cite{mannucci2010}, \cite{andrews2013} and \cite{curti2020}. We get particularly good agreement with the results derived by \cite{andrews2013} and \cite{curti2020}. The metallicity measurements for both these works are (anchored on) ``direct'' metallicity measurements, which means that we can perform a one-to-one comparison with our measurements. Below $\sim 10^{8.5}$ \msun we are seeing some deviations with the result from \cite{andrews2013}, however, at these stellar masses the SDSS sample has much lower completeness than DESI. The discrepancies with the \citep{tremonti2004, pettini2004} and \cite{mannucci2010} calibrations are due to differences in metallicity calibrations which are well documented \citep[e.g.][]{kewley2008}.

\begin{figure*}
    \begin{center}
    \begin{subfigure}{0.495\textwidth}
        \includegraphics[width=\textwidth]{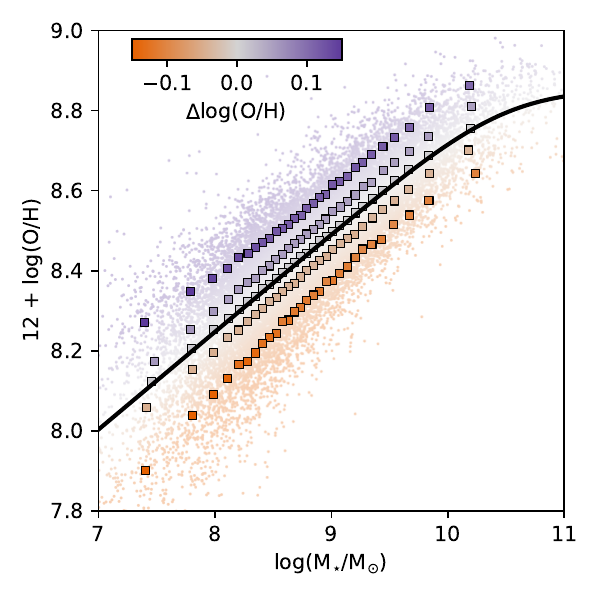}
    \end{subfigure}
    \hfill
    \begin{subfigure}{0.495\textwidth}
        \includegraphics[width=\textwidth]{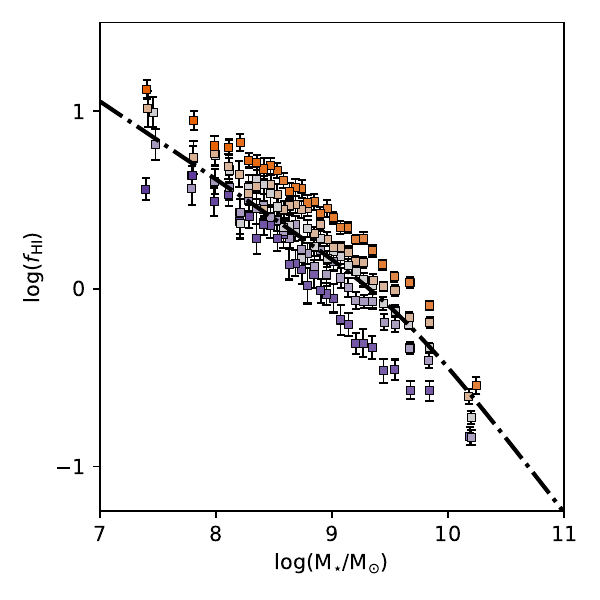}
    \end{subfigure}
    \caption{\textit{Left}: The mass-metallicity relation binned in stellar mass and offset from the average MZR ($\Delta$log(O/H), black, solid, same as in Fig. \ref{fig:mzr}). For each bin the average stellar mass and metallicity is shown (square) coloured by offset from the mass-metallicity relation. \textit{Right}: The atomic gas sequence (black, dash-dotted, same as in Fig. \ref{fig:fhi_vs_mstar}) and measured atomic gas fraction of the galaxies in each stellar mass/MZR offset bin. The binned measurements are coloured by the offset from the MZR as in the left panel.}
    \label{fig:scatter_mzr_fhi}
    \end{center}
\end{figure*}

\subsubsection{The mass-metallicity relation for a representative sample of star forming galaxies}
We repeat the metallicity measurements for a more representative sample of star forming galaxies based on our mass complete sample. Only galaxies with some star formation produce the emission lines used to measure the gas phase metallicity. Therefore, we exclude quiescent galaxies from our sample by introducing a S/N > 3 selection on the H-alpha flux; below a stellar mass of $10^9~\textrm{M}_{\odot}$ this excludes only $\sim 9$\% of our sample, above a stellar mass of $10^9~\textrm{M}_{\odot}$ this increases to $\sim 15$\%. As only a small fraction of the mass complete sample is excluded through this selection we make comparisons between the metallicity measurements of this representative sample of star forming galaxies and gas fraction measurements of the mass complete sample; the exclusion of this small subset of galaxies does not have a large influence on the median measurement of the atomic gas fraction as we show in Table \ref{tab:atomic_gas_measurements} in Appendix \ref{sec:appendix_a}. As we do not have reliable emission line measurements for the individual galaxies in this sample we can not remove AGN. This population contaminates the sample at high stellar mass and therefore we are only able to make reliable metallicity measurements from the mass-complete stacks at $M_{\star} < 10^{10.3} \textrm{M}_{\odot}$ \citep[e.g.][]{kauffmann2003}. For this mass-complete sample, we use spectral stacking to ensure all the required emission lines are detected. We divide the sample into stellar mass bins, each containing $\sim 50$ galaxies. We show the results from our stacked measurements in the right panel of Figure \ref{fig:mzr}. These results for what is a much more complete sample show a significantly shallower slope of the MZR at masses $<10^9 \textrm{M}_{\odot}$, compared to the emission line flux limited sample. The difference between these samples is mainly the inclusion in the mass-complete sample of galaxies with a lower star formation rate and gas fraction. As shown by the FMR, at fixed stellar mass, metallicity is inversely correlated with star formation rate, therefore the inclusion of galaxies with a low star formation rate will increase the average gas phase metallicity of the sample \citep[e.g.][]{ellison2008, mannucci2010}. This is also reflected in the lower average gas fraction of the atomic gas sequence of the mass complete sample compared to the emission line flux limited sample as shown in Figure \ref{fig:fhi_vs_mstar}.

\subsection{The connection between the scatter in the mass-metallicity relation and atomic gas sequence}
Our results show that sample selection has a significant effect on the measured atomic gas sequence and mass-metallicity relation. At fixed stellar mass, the gas fraction/metallicity of galaxies is systematically higher/lower for the emission line limited sample compared to the mass complete sample. This shows that the atomic gas mass and metallicity of galaxies are strongly related quantities.

We investigate further this connection between the mass-metallicity relation and the atomic gas sequence, by looking for systematic trends in the scatter around the scaling relations. To do so, we divide the galaxies in our emission line flux limited sample and ALFALFA coverage into 34 quantile bins in stellar mass and 5 quantile bins in offset from the average mass-metallicity relation (Eq. \ref{eq:mzr_curti}); this ensures that each bin contains $\sim 150$ galaxies. The average stellar mass and metallicity of the galaxies in each bin is shown in the left panel of Figure \ref{fig:scatter_mzr_fhi}. We stack the ALFALFA spectra in these stellar mass/MZR offset bins to measure the atomic gas mass in each bin; this is shown in the right panel of Figure \ref{fig:scatter_mzr_fhi}. Our measurements show that for all stellar mass bins, the most metal-poor(rich) galaxies are also the most gas-rich(poor). This result is in agreement with several previous studies which have revealed that gas mass plays an important role in establishing the scatter of the mass-metallicity relation \citep{brinchmann2013, bothwell2013, bothwell2016, lara-lopez2013, hughes2013, brown2018, scholte2023}. In this work we push these established relations to much lower stellar masses ($10^{7.3} ~\textrm{M}_{\odot}$) than most previous studies ($\sim10^9 ~\textrm{M}_{\odot}$). Qualitatively, we get similar results to the $\rm FMR_{HI}$ measurements from \cite{jimmy2015} in the dwarf galaxy regime, however, here derived using a much larger sample. 

\section{Discussion}
\label{sec:discussion}
Using the DESI and ALFALFA surveys we constrain the atomic gas sequence and mass-metallicity relation over a stellar mass range from $10^{6.5}~\textrm{M}_{\odot}$ to $10^{11.5}~\textrm{M}_{\odot}$. We find that there is a slope change in both the atomic gas sequence and the MZR at $\sim 10^{9}~\textrm{M}_{\odot}$, when using a mass-complete sample. We also find that the scatter of the mass-metallicity relation is inversely correlated with the scatter of the atomic gas sequence over the entire mass range. These measurements of the slope change in both the atomic gas sequence and MZR at a similar mass of $\sim10^9 ~\textrm{M}_{\odot}$ (see Fig. \ref{fig:fhi_vs_mstar} and the right-hand panel of Fig. \ref{fig:mzr}) provide compelling evidence that there is a transition of the dominant process that regulates the evolution of dwarf galaxies and massive galaxies, respectively.

\subsection{A transition of the atomic gas sequence at $\rm \sim 10^{9} ~\textrm{M}_{\odot}$}
The measurement of a slope change in the atomic gas sequence indicates a change in the dominant processes that sets the atomic gas mass between the dwarf and massive galaxy regimes. The combination of efficient accretion of cool gas onto low mass galaxies \citep["cold mode accretion", e.g.][]{keres2005}, and efficient removal of gas from the shallow potential well of dwarfs due to supernova feedback are likely to be dominant processes determining the gas content of dwarf galaxies \citep[e.g.][]{larson1974}. 

\cite{dekel1986} predicted that supernova-driven winds efficiently drive a substantial fraction of the gas out of galaxies when the virial velocity is lower than a critical value ($\rm V_{crit}$) of $\rm \sim 100 ~km$ $\rm s^{-1}$. More recently, simulations with physically motivated supernova feedback models also show a threshold value of the maximum rotational velocity of $\rm \sim 100 ~km$ $\rm s^{-1}$ allowing the removal of a significant fraction of the gas phase of dwarfs \citep{derossi2010}. The supernova feedback model in these simulations is able to explain the observed break in the stellar Tully-Fisher relation at $\rm \sim 90 ~km$ $\rm s^{-1}$ \citep{tully1977, mcgaugh2000}. Other simulations have also indicated that there is significant gas removal from dwarfs through feedback; the ISM of dwarf galaxies has a much lower density and is not collapsed into a thin disk as observed at higher masses \citep[e.g.][]{hopkins2012, shen2014, rey2022}. This may make the gas phase of dwarf galaxies much more susceptible to feedback. \cite{hopkins2012} have shown that in the dwarf galaxy regime a combination of momentum from UV photons, stellar winds, warm gas pressure from photoionization regions and supernovae feedback can generate a multiphase outflow which ejects large amounts of gas.  The literature results above also show that this process ceases to be efficient at masses above $\rm \sim 10^{9} ~\textrm{M}_{\odot}$. 

Measurements of the outflow velocities ($\rm V_{out}$) in star forming galaxies at $\rm 10^{9} ~\textrm{M}_{\odot}$ show that $\rm V_{out} \simeq 100 ~km$ $\rm s^{-1}$ $\rm \simeq V_{crit}$ and that at higher masses $\rm V_{out} < V_{crit}$ \citep{roberts-borsani2019}. This shows that the outflow velocities of normal star forming galaxies are not high enough to escape the potential.  However, in the shallower potentials of dwarfs, outflows can escape into the intergalactic medium. With decreasing mass, galaxies are significantly more efficient at generating outflows; simulations show us that this is largely driven by the effect of the halo potential well \citep{christensen2016}.

\begin{figure}
    \begin{center}
    \includegraphics[width=0.5\textwidth]{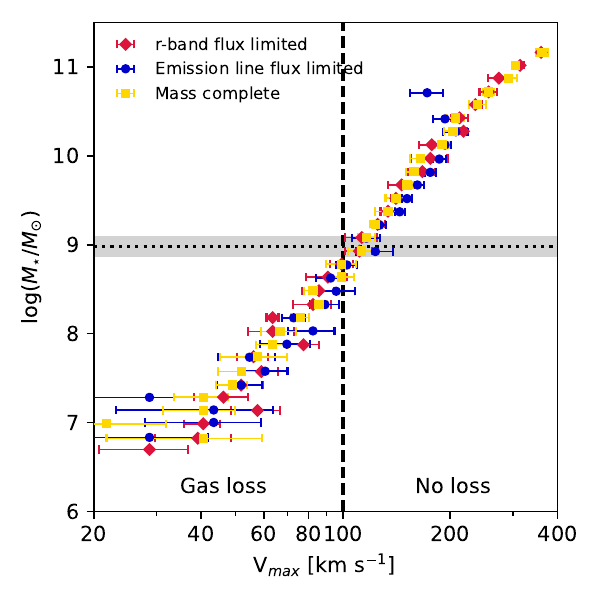} 
    \caption{The maximum rotational velocity versus stellar mass of galaxies \protect\citep[stellar Tully-Fisher relation;][]{tully1977}. The data points show the measured maximum rotational velocity inferred from the $\rm W_{20}$ line width of the 21-cm line (see Section \ref{sec:methods_alfalfa}). The vertical dashed line shows the critical value of the maximum rotational velocity below which gas loss is expected according to \protect\cite{dekel1986}. The horizontal dotted line and grey band denotes the location of the slope transition of the atomic gas sequence in our mass complete sample: $\log(M_{\star}/\textrm{M}_{\odot}) = 8.98_{-0.11}^{+0.11}$.}
    \label{fig:vmax_mstar}
    \end{center}
\end{figure}

We can estimate the characteristic stellar mass at which galaxies transition into the regime where significant gas loss is expected. To do this, we calculate the maximum rotational velocity of the galaxies in our sample through the width of the 21-cm emission line in the stacked ALFALFA spectra. The result is displayed in Figure \ref{fig:vmax_mstar}, which shows that the expected transition at a virial velocity of $\rm \sim 100 ~km$ $\rm s^{-1}$ indeed corresponds to a stellar mass of $\rm \sim 10^{9} ~\textrm{M}_{\odot}$. Based on this, we can infer that the change in efficiency of SNe-driven outflows in galaxies with virial velocities greater/smaller than 100 km~s$^{-1}$ is likely to be a contributing factor to the slope change of the atomic gas sequence observed at $\rm \sim 10^{9} ~\textrm{M}_{\odot}$

At stellar masses larger than $\rm \sim 10^{9} ~\textrm{M}_{\odot}$, the relation between the stellar mass and HI gas content of galaxies is unlikely to be regulated in most part by supernova feedback. In this higher mass regime, we see a persistent trend where the gas fraction of galaxies decreases with stellar mass, as previously also robustly derived by the xGASS survey \citep{catinella2018}. If gas removal is unlikely to be the dominant factor regulating this steep declining slope at masses $>10^{9} ~\textrm{M}_{\odot}$ (at least until significantly higher masses where AGN feedback might play a role), then we should logically look instead for changes in gas accretion. 

It has been convincingly argued based on numerical simulations that there is a change in the efficiency of gas accretion onto galaxies at stellar masses of $10^{10}~\textrm{M}_{\odot}$ \citep[e.g.][]{keres2005, dekel2006, vandevoort2011}.  Above such a characteristic mass infalling gas is shock heated to the virial temperature of the halo. The very long cooling time of shock heated gas makes the accretion of this gas into the ISM an inefficient process, effectively starving the galaxy of fresh gas.  This naturally leads to a relative decrease of the total cold gas reservoirs of galaxies as they grow more massive. While the transition between the so-called ``cold-mode" and ``hot-mode" accretion onto halos is often reported to occur at stellar masses of $\sim10^{10}~\textrm{M}_{\odot}$, a smoother transition between the two modes could explain why the gas reservoirs (relative to the stellar mass) are already starting to decline at $\sim10^9~\textrm{M}_{\odot}$.  
A thorough analysis of the respective roles of gas inflows and outflows in determining the slope of the atomic gas sequence will in the future be enabled through detailed comparisons with modern simulations. These simulations are starting to reach the volumes and resolution required to produce galaxy scaling relations over similarly broad stellar mass ranges \citep[e.g.][]{feldmann2023}. 

\subsection{Understanding the link between the atomic gas sequence and the MZR through a simple analytical model}

\begin{figure*}
    \begin{center}
    \begin{subfigure}{0.495\textwidth}
        \includegraphics[width=\textwidth]{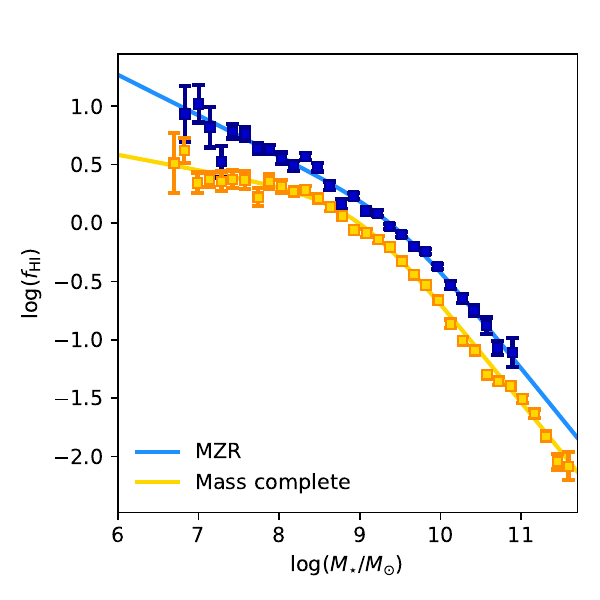}
    \end{subfigure}
    \hfill
    \begin{subfigure}{0.495\textwidth}
        \includegraphics[width=\textwidth]{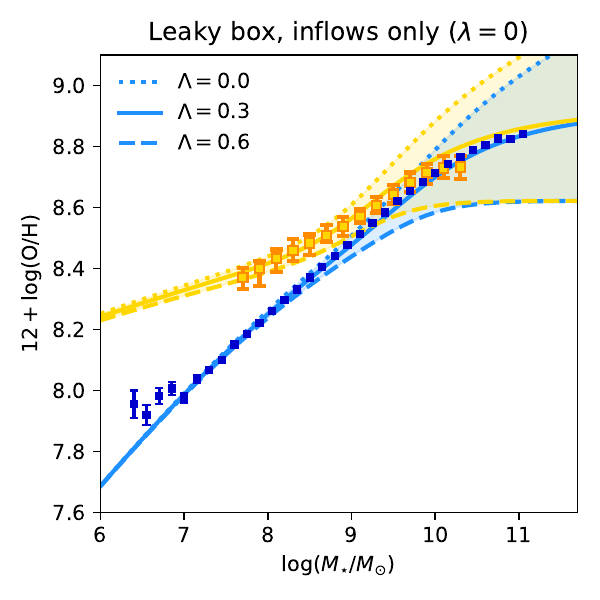}
    \end{subfigure}
    \hfill
    \begin{subfigure}{0.495\textwidth}
        \includegraphics[width=\textwidth]{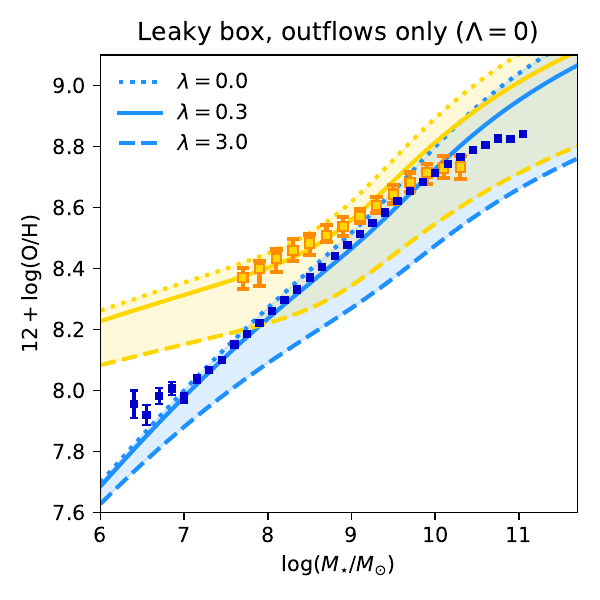}
    \end{subfigure}
    \hfill
    \begin{subfigure}{0.495\textwidth}
        \includegraphics[width=\textwidth]{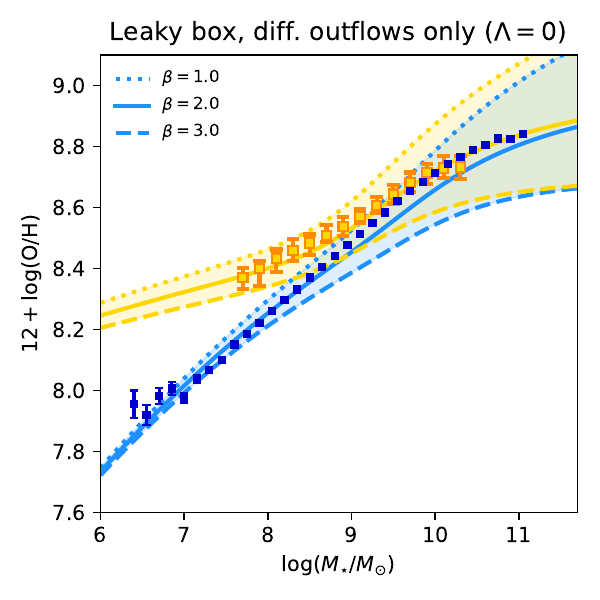}
    \end{subfigure}
    \caption{Modelled mass-metallicity relations derived using the analytical model from \protect\cite{recchi2008}. \textit{Top left:} The measured atomic gas sequence as in Fig. \ref{eq:atomic_gas_sequence} from the emission line limited sample (dark/light blue) together with the atomic gas sequence of the mass complete sample (orange/yellow). \textit{Top right:} The measured MZRs (black, same as in Fig. \ref{fig:mzr}) for the line flux limited (dark blue) and mass complete (orange) samples shown together with MZRs derived using the analytical model using the respective gas sequences coloured according to the top-left figure. These MZRs are derived assuming that the metallicity is regulated by inflows of pristine gas and no outflows. Line styles show different values of the inflow parameter strength as explained by the legend. \textit{Bottom left:} Same as top right except here showing the results where the metallicity is regulated by outflows. \textit{Bottom right:} Same as top right except here showing the results where the metallicity is regulated by differential outflows.}
    \label{fig:analytic_model}
    \end{center}	
\end{figure*}

The shape and scatter of the MZR is often attributed to the effects of inflows of pristine gas and removal of metals from the gas reservoirs of galaxies through feedback. To gain an understanding of how these processes may connect to explain our results on the atomic gas sequence and MZR, we use a simple chemical evolution model \citep{recchi2008}. This model describes the ISM of galaxies as a gas reservoir in which metals are deposited through star formation. The reservoir can be fueled by (metal poor) inflows and metals can be removed through (metal rich) outflows. The gas phase metallicity is determined using the following equation:
\begin{equation}
\begin{split}
    Z = & \frac{\Lambda Z_{A} + y_{Z}}{\Lambda + (\beta - 1)\lambda} \times \\ & \left( 1 - \left[ (\Lambda - \lambda) - (\Lambda - \lambda - 1)\mu^{-1} \right]^{\frac{\Lambda+(\beta - 1)\lambda}{\Lambda -\lambda - 1}} \right)
\end{split}
\end{equation}
where $y_{Z}$ is the yield of metals through star formation, $\mu$ is the gas fraction $M_{\textrm{gas}}/(M_{\star}+M_{\textrm{gas}})$, $\Lambda$ and $\lambda$ are the relative strengths of gas inflows and feedback proportional to the star formation rate, $Z_{A}$ is the metallicity of inflowing gas and $\beta$ regulates the preferential ejection of metal enriched gas. Although simplistic and not encompassing all of the complexity of the underlying physical processes, this kind of simple analytical models are quite successful at reproducing the MZR \citep[e.g.][]{lilly2013,recchi2008,tortora22}. Here we use the model to provide us with a qualitative understanding of how different processes affect the shape of the MZR, rather than to try to put quantitative constraints on quantities such as mass loading factors, metal yields, or IMF variations due to strong degeneracies between model parameters \citep[e.g.][]{spitoni10}. The behaviour seen in the analytic model used here can be reproduced using similar analytical prescriptions \citep[e.g.][]{dave2012, lilly2013}.

We use our measurements of the atomic gas sequence to determine $\mu$ and set the following parameters to fixed values for the inflow and outflow models: $y_{Z} = 0.0084$, $Z_{A} = 0.0$. In the differential outflow model we fix the parameters to: $y_{Z} = 0.010$, $Z_{A} = 0.0$, $\lambda = 0.3$. In these models $\mu$ represents the gas fraction that is actively involved in the star forming process and in which the metals are mixed $\sim$instantly. This is not the case for the full atomic gas mass, therefore we assume that 25\% of the gas mass is "star forming gas", which roughly corresponds to the atomic to molecular gas ratio \citep[e.g.][]{saintonge2011, catinella2018}. 

In the top-left panel of Figure \ref{fig:analytic_model} we show the measured atomic gas sequence as in for the line flux limited sample (blue) as well as the mass complete sample (orange). The remaining panels show the expected MZR using models including outflows (top-right), inflows (bottom-left) and differential outflows (bottom-right), each using the two different atomic gas sequences as input. For each model we show several values for the inflow ($\Lambda$), outflow ($\lambda$) and differential outflow ($\beta$) parameters, as shown in the legends of the respective graphs. These panels show clearly that the shape of the atomic gas sequence is the main factor responsible for the slope of the MZR at masses $<10^{9}~\textrm{M}_{\odot}$. Models with the same parameter values for inflows and outflows can simultaneously describe the MZR slopes of both of these samples, as long as the relevant atomic gas sequence is used as input. 

\section{Conclusions}
\label{sec:conclusions}
Using combined observations from the DESI and ALFALFA surveys we constrained the atomic gas sequence and mass-metallicity relation from dwarfs ($10^{6.5} ~\textrm{M}_{\odot}$) to massive galaxies ($10^{11.5} ~\textrm{M}_{\odot}$). This is made possible due to the faint magnitude limit of the DESI survey in comparison with previous generations of large spectroscopic surveys such as SDSS. Summarised our main findings are:
\begin{itemize}
    \item We find a slope change of the atomic gas sequence at $\sim 10^{9}~ \textrm{M}_{\odot}$. In the dwarf galaxy regime the slope of the atomic gas sequence becomes shallower. We show that the mass range at which we see a shallow slope of atomic gas sequence change is consistent with predictions of supernova driven gas loss in the dwarf galaxy regime \citep[e.g.][]{larson1974, dekel1986, derossi2010}.
    \item We measure the mass-metallicity relation using individual galaxies which is characterised by a high mass plateau at $\sim 10^{10.5} ~\textrm{M}_{\odot}$ and a power-law relation with a constant slope of $\sim 0.24$ at lower mass. Our results are in excellent agreement with other works using similar metallicity measurements \citep[e.g.][]{andrews2013, curti2020}. However, the faint magnitude limit of the DESI survey allows us to constrain the MZR down to galaxies with $M_{\star} \sim 10^{6.5} ~\textrm{M}_{\odot}$.
    \item We show that the slope of the mass-metallicity relation is however significantly shallower at stellar masses $< 10^{9}~ \textrm{M}_{\odot}$, when measured for a much more complete and representative sample of star forming galaxies. Using the chemical evolution model of \cite{recchi2008} to argue that the slope change of the MZR at $\sim 10^{9}~ \textrm{M}_{\odot}$ is directly linked to the behaviour of the atomic gas reservoir. 
\end{itemize}

The joint constraints on these scaling relations over 5 orders of magnitude in stellar mass provide compelling evidence that there is a transition of the dominant processes that regulate the evolution of dwarf galaxies and massive galaxies, respectively.  Further DESI data releases will significantly increase the number of low mass galaxies and the overlap with the ALFALFA survey, allowing us to better constrain the joint behaviour  of galaxy scaling relations, and to interpret the observations through a combination of analytic models and large volume simulations. 

\section*{Acknowledgements}
The authors thank the DESI publication board, in particular Segev BenZvi, for their part in improving the quality of this work.

Parts of this research were supported by the Australian Research Council Centre of Excellence for All Sky Astrophysics in 3 Dimensions (ASTRO 3D), through project number CE170100013.

H.Z. acknowledges the support from the National Key R\&D Program of China (grant Nos. 2022YFA1602902 and 2023YFA1607800) and the National Natural Science Foundation of China (NSFC; grant Nos. 12120101003 and 12373010). 

This material is based upon work supported by the U.S. Department of Energy (DOE), Office of Science, Office of High-Energy Physics, under Contract No. DE–AC02–05CH11231, and by the National Energy Research Scientific Computing Center, a DOE Office of Science User Facility under the same contract. Additional support for DESI was provided by the U.S. National Science Foundation (NSF), Division of Astronomical Sciences under Contract No. AST-0950945 to the NSF’s National Optical-Infrared Astronomy Research Laboratory; the Science and Technology Facilities Council of the United Kingdom; the Gordon and Betty Moore Foundation; the Heising-Simons Foundation; the French Alternative Energies and Atomic Energy Commission (CEA); the National Council of Science and Technology of Mexico (CONACYT); the Ministry of Science and Innovation of Spain (MICINN), and by the DESI Member Institutions: \url{https://www.desi.lbl.gov/collaborating-institutions}. Any opinions, findings, and conclusions or recommendations expressed in this material are those of the author(s) and do not necessarily reflect the views of the U. S. National Science Foundation, the U. S. Department of Energy, or any of the listed funding agencies.

The authors are honored to be permitted to conduct scientific research on Iolkam Du’ag (Kitt Peak), a mountain with particular significance to the Tohono O’odham Nation.

The DESI Legacy Imaging Surveys consist of three individual and complementary projects: the Dark Energy Camera Legacy Survey (DECaLS), the Beijing-Arizona Sky Survey (BASS), and the Mayall z-band Legacy Survey (MzLS). DECaLS, BASS and MzLS together include data obtained, respectively, at the Blanco telescope, Cerro Tololo Inter-American Observatory, NSF’s NOIRLab; the Bok telescope, Steward Observatory, University of Arizona; and the Mayall telescope, Kitt Peak National Observatory, NOIRLab. NOIRLab is operated by the Association of Universities for Research in Astronomy (AURA) under a cooperative agreement with the National Science Foundation. Pipeline processing and analyses of the data were supported by NOIRLab and the Lawrence Berkeley National Laboratory (LBNL). Legacy Surveys also uses data products from the Near-Earth Object Wide-field Infrared Survey Explorer (NEOWISE), a project of the Jet Propulsion Laboratory/California Institute of Technology, funded by the National Aeronautics and Space Administration. Legacy Surveys was supported by: the Director, Office of Science, Office of High Energy Physics of the U.S. Department of Energy; the National Energy Research Scientific Computing Center, a DOE Office of Science User Facility; the U.S. National Science Foundation, Division of Astronomical Sciences; the National Astronomical Observatories of China, the Chinese Academy of Sciences and the Chinese National Natural Science Foundation. LBNL is managed by the Regents of the University of California under contract to the U.S. Department of Energy. The complete acknowledgments can be found at \url{https://www.legacysurvey.org/acknowledgment/}.

\section*{Data Availability}
The data from the DESI Survey Validation period is publicly available at \url{https://data.desi.lbl.gov/doc/} \citep{desi2023}. The full DESI Survey Year 1 dataset used in this work will be made public at the same location in 2025. This includes spectra and derived data such as emission line flux measurements from \textsc{FastSpecFit} \citep{moustakas2023}. The Legacy Survey imaging is available at \url{https://www.legacysurvey.org/}. Data products from the ALFALFA Survey are available at \url{https://egg.astro.cornell.edu/index.php/}. Other derived data generated in this research will be shared on reasonable request to the corresponding author.



\bibliographystyle{mnras}
\bibliography{main_bib} 




\appendix

\section{Measurements of the atomic gas sequence and mass metallicity relation}
\label{sec:appendix_a}
This appendix contains tables of the measured atomic gas sequences and mass-metallicity relations for the samples used in this paper. The atomic gas sequence measurements are shown in Table \ref{tab:atomic_gas_measurements}. The measurements of the mass-metallicity relation derived from individual galaxies in the emission line flux limited sample are shown in Table \ref{tab:mzr_measurements}. The measurements of the mass-metallicity relation derived from stacked spectra of the representative sample of star forming galaxies are shown in Table \ref{tab:stacked_mzr_measurements}.

\begin{table*}
	\caption{The atomic gas fraction measurements of the samples used in this paper.}
	\label{tab:atomic_gas_measurements}
\centering
\begin{tabular}{ccccc}
\hline \hline
\textbf{$\log(M_{\star}/\textrm{M}_\odot)$} & \multicolumn{4}{c}{\textbf{$\log(F_{\textrm{HI}})$}} \\ \cline{2-5}
 & Flux limited & Line flux limited & Mass complete & Representative SF \\ \hline 
$  6.675 $ & $ 0.75 \pm 0.09 $ & --- & $ 0.5 \pm 0.3 $ & --- \\
$  6.825 $ & $ 0.82 \pm 0.09 $ & $ 0.94 \pm 0.24 $ & $ 0.62 \pm 0.11 $ & $ 0.70 \pm 0.08 $ \\
$  6.975 $ & $ 0.85 \pm 0.09 $ & $ 1.02 \pm 0.16 $ & $ 0.34 \pm 0.09 $ & $ 0.35 \pm 0.10 $ \\
$  7.125 $ & $ 0.82 \pm 0.06 $ & $ 0.82 \pm 0.17 $ & $ 0.37 \pm 0.06 $ & $ 0.38 \pm 0.07 $ \\
$  7.275 $ & $ 0.71 \pm 0.06 $ & $ 0.53 \pm 0.14 $ & $ 0.36 \pm 0.09 $ & $ 0.39 \pm 0.08 $ \\
$  7.425 $ & $ 0.74 \pm 0.04 $ & $ 0.79 \pm 0.06 $ & $ 0.37 \pm 0.08 $ & $ 0.39 \pm 0.07 $ \\
$  7.575 $ & $ 0.79 \pm 0.04 $ & $ 0.76 \pm 0.06 $ & $ 0.37 \pm 0.07 $ & $ 0.40 \pm 0.07 $ \\
$  7.725 $ & $ 0.62 \pm 0.04 $ & $ 0.64 \pm 0.05 $ & $ 0.22 \pm 0.08 $ & $ 0.28 \pm 0.08 $ \\
$  7.875 $ & $ 0.68 \pm 0.04 $ & $ 0.63 \pm 0.04 $ & $ 0.35 \pm 0.06 $ & $ 0.35 \pm 0.06 $ \\
$  8.025 $ & $ 0.65 \pm 0.04 $ & $ 0.55 \pm 0.05 $ & $ 0.32 \pm 0.05 $ & $ 0.34 \pm 0.05 $ \\
$  8.175 $ & $ 0.54 \pm 0.04 $ & $ 0.49 \pm 0.04 $ & $ 0.27 \pm 0.04 $ & $ 0.31 \pm 0.04 $ \\
$  8.325 $ & $ 0.47 \pm 0.04 $ & $ 0.57 \pm 0.03 $ & $ 0.28 \pm 0.03 $ & $ 0.31 \pm 0.03 $ \\
$  8.475 $ & $ 0.45 \pm 0.04 $ & $ 0.48 \pm 0.04 $ & $ 0.21 \pm 0.04 $ & $ 0.29 \pm 0.03 $ \\
$  8.625 $ & $ 0.34 \pm 0.04 $ & $ 0.32 \pm 0.04 $ & $ 0.14 \pm 0.03 $ & $ 0.19 \pm 0.03 $ \\
$  8.775 $ & $ 0.17 \pm 0.04 $ & $ 0.16 \pm 0.04 $ & $ 0.06 \pm 0.03 $ & $ 0.16 \pm 0.03 $ \\
$  8.925 $ & $ 0.03 \pm 0.04 $ & $ 0.23 \pm 0.03 $ & $ -0.06 \pm 0.04 $ & $ 0.04 \pm 0.03 $ \\
$  9.075 $ & $ -0.01 \pm 0.04 $ & $ 0.10 \pm 0.03 $ & $ -0.09 \pm 0.04 $ & $ 0.045 \pm 0.024 $ \\
$  9.225 $ & $ -0.09 \pm 0.03 $ & $ 0.08 \pm 0.03 $ & $ -0.14 \pm 0.03 $ & $ -0.016 \pm 0.023 $ \\
$  9.375 $ & $ -0.28 \pm 0.04 $ & $ -0.028 \pm 0.023 $ & $ -0.21 \pm 0.03 $ & $ -0.09 \pm 0.03 $ \\
$  9.525 $ & $ -0.33 \pm 0.03 $ & $ -0.097 \pm 0.023 $ & $ -0.33 \pm 0.03 $ & $ -0.23 \pm 0.03 $ \\
$  9.675 $ & $ -0.43 \pm 0.03 $ & $ -0.201 \pm 0.022 $ & $ -0.44 \pm 0.03 $ & $ -0.35 \pm 0.03 $ \\
$  9.825 $ & $ -0.51 \pm 0.03 $ & $ -0.245 \pm 0.022 $ & $ -0.53 \pm 0.03 $ & $ -0.39 \pm 0.03 $ \\
$  9.975 $ & $ -0.70 \pm 0.04 $ & $ -0.375 \pm 0.021 $ & $ -0.66 \pm 0.03 $ & $ -0.56 \pm 0.03 $ \\
$  10.125 $ & $ -0.91 \pm 0.04 $ & $ -0.53 \pm 0.03 $ & $ -0.86 \pm 0.04 $ & $ -0.70 \pm 0.03 $ \\
$  10.275 $ & $ -1.00 \pm 0.04 $ & $ -0.65 \pm 0.04 $ & $ -1.01 \pm 0.03 $ & $ -0.84 \pm 0.03 $ \\
$  10.425 $ & $ -1.10 \pm 0.03 $ & $ -0.75 \pm 0.05 $ & $ -1.09 \pm 0.04 $ & $ -0.90 \pm 0.03 $ \\
$  10.575 $ & $ -1.31 \pm 0.04 $ & $ -0.88 \pm 0.07 $ & $ -1.30 \pm 0.03 $ & $ -1.11 \pm 0.03 $ \\
$  10.725 $ & $ -1.33 \pm 0.04 $ & $ -1.07 \pm 0.06 $ & $ -1.36 \pm 0.04 $ & $ -1.18 \pm 0.03 $ \\
$  10.875 $ & $ -1.50 \pm 0.04 $ & $ -1.11 \pm 0.12 $ & $ -1.40 \pm 0.03 $ & $ -1.31 \pm 0.03 $ \\
$  11.025 $ & $ -1.51 \pm 0.03 $ & --- & $ -1.51 \pm 0.03 $ & $ -1.43 \pm 0.03 $ \\
$  11.175 $ & $ -1.64 \pm 0.04 $ & --- & $ -1.63 \pm 0.04 $ & $ -1.53 \pm 0.03 $ \\
$  11.325 $ & $ -1.82 \pm 0.04 $ & --- & $ -1.83 \pm 0.04 $ & $ -1.71 \pm 0.04 $ \\
$  11.475 $ & $ -2.03 \pm 0.06 $ & --- & $ -2.05 \pm 0.07 $ & $ -1.93 \pm 0.07 $ \\
$  11.625 $ & $ -2.08 \pm 0.12 $ & --- & $ -2.08 \pm 0.12 $ & $ -1.88 \pm 0.12 $ \\
$  11.775 $ & $ -2.3 \pm 0.4 $ & --- & $ -2.3 \pm 0.4 $ & --- \\ \hline
\end{tabular}
\end{table*}

\begin{table}
	\caption{The metallicity measurements of the emission line flux limited sample.}
	\label{tab:mzr_measurements}
\centering
\begin{tabular}{rrrr}
\hline \hline
\textbf{$\log(M_{\star}/\textrm{M}_\odot)$} & \multicolumn{3}{c}{\textbf{$12 + \log(\textrm{O/H})$}} \\ \cline{2-4}
 & Mean & Standard deviation & Standard error \\ \hline 
$  6.40 $ & $ 7.955 $ & $ 0.252 $ & $ 0.0453 $ \\
$  6.55 $ & $ 7.919 $ & $ 0.241 $ & $ 0.0325 $ \\
$  6.70 $ & $ 7.983 $ & $ 0.218 $ & $ 0.0255 $ \\
$  6.85 $ & $ 8.007 $ & $ 0.218 $ & $ 0.0210 $ \\
$  7.00 $ & $ 7.974 $ & $ 0.207 $ & $ 0.0159 $ \\
$  7.15 $ & $ 8.037 $ & $ 0.217 $ & $ 0.0133 $ \\
$  7.30 $ & $ 8.068 $ & $ 0.199 $ & $ 0.0099 $ \\
$  7.45 $ & $ 8.100 $ & $ 0.179 $ & $ 0.0076 $ \\
$  7.60 $ & $ 8.151 $ & $ 0.171 $ & $ 0.0058 $ \\
$  7.75 $ & $ 8.185 $ & $ 0.158 $ & $ 0.0044 $ \\
$  7.90 $ & $ 8.221 $ & $ 0.144 $ & $ 0.0033 $ \\
$  8.05 $ & $ 8.262 $ & $ 0.129 $ & $ 0.0024 $ \\
$  8.20 $ & $ 8.297 $ & $ 0.120 $ & $ 0.0018 $ \\
$  8.35 $ & $ 8.331 $ & $ 0.113 $ & $ 0.0015 $ \\
$  8.50 $ & $ 8.370 $ & $ 0.102 $ & $ 0.0012 $ \\
$  8.65 $ & $ 8.406 $ & $ 0.095 $ & $ 0.0011 $ \\
$  8.80 $ & $ 8.441 $ & $ 0.089 $ & $ 0.0010 $ \\
$  8.95 $ & $ 8.477 $ & $ 0.087 $ & $ 0.0010 $ \\
$  9.10 $ & $ 8.513 $ & $ 0.085 $ & $ 0.0010 $ \\
$  9.25 $ & $ 8.550 $ & $ 0.084 $ & $ 0.0011 $ \\
$  9.40 $ & $ 8.584 $ & $ 0.082 $ & $ 0.0012 $ \\
$  9.55 $ & $ 8.621 $ & $ 0.080 $ & $ 0.0013 $ \\
$  9.70 $ & $ 8.654 $ & $ 0.080 $ & $ 0.0015 $ \\
$  9.85 $ & $ 8.684 $ & $ 0.076 $ & $ 0.0017 $ \\
$  10.00 $ & $ 8.714 $ & $ 0.080 $ & $ 0.0023 $ \\
$  10.15 $ & $ 8.742 $ & $ 0.080 $ & $ 0.0030 $ \\
$  10.30 $ & $ 8.766 $ & $ 0.078 $ & $ 0.0034 $ \\
$  10.45 $ & $ 8.788 $ & $ 0.091 $ & $ 0.0050 $ \\
$  10.60 $ & $ 8.805 $ & $ 0.073 $ & $ 0.0053 $ \\
$  10.75 $ & $ 8.827 $ & $ 0.069 $ & $ 0.0058 $ \\
$  10.90 $ & $ 8.824 $ & $ 0.059 $ & $ 0.0070 $ \\
$  11.05 $ & $ 8.841 $ & $ 0.049 $ & $ 0.0093 $ \\ \hline
\end{tabular}
\end{table}

\begin{table}
	\caption{The metallicity measurements of the representative sample of SF galaxies.}
	\label{tab:stacked_mzr_measurements}
\centering
\begin{tabular}{rr}
\hline \hline
\textbf{$\log(M_{\star}/\textrm{M}_\odot)$} & \textbf{$12 + \log(\textrm{O/H})$} \\ \hline
$  7.8 $ & $ 8.37 _{- 0.04 }^{+ 0.03 }$ \\
$  8.0 $ & $ 8.40 _{- 0.06 }^{+ 0.03 }$ \\
$  8.2 $ & $ 8.43 _{- 0.05 }^{+ 0.03 }$ \\
$  8.4 $ & $ 8.46 _{- 0.04 }^{+ 0.04 }$ \\
$  8.6 $ & $ 8.48 _{- 0.04 }^{+ 0.03 }$ \\
$  8.8 $ & $ 8.508 _{- 0.021 }^{+ 0.035 }$ \\
$  9.0 $ & $ 8.54 _{- 0.03 }^{+ 0.03 }$ \\
$  9.2 $ & $ 8.570 _{- 0.020 }^{+ 0.026 }$ \\
$  9.4 $ & $ 8.61 _{- 0.03 }^{+ 0.03 }$ \\
$  9.6 $ & $ 8.64 _{- 0.03 }^{+ 0.03 }$ \\
$  9.8 $ & $ 8.68 _{- 0.03 }^{+ 0.03 }$ \\
$  10.0 $ & $ 8.71 _{- 0.04 }^{+ 0.03 }$ \\
$  10.2 $ & $ 8.73 _{- 0.03 }^{+ 0.04 }$ \\ \hline
\end{tabular}
\end{table}
\section{Author affiliations}
\textit{
$^1$Department of Physics \& Astronomy, University College London, Gower Street, London, WC1E 6BT, UK\\
$^{2}$Institute for Astronomy, University of Edinburgh, Royal Observatory, Edinburgh, EH9 3HJ, UK\\
$^{3}$Department of Physics and Astronomy, Siena College, 515 Loudon Road, Loudonville, NY 12110, USA\\
$^{4}$International Centre for Radio Astronomy Research, The University of Western Australia, Crawley, WA 6009, Australia\\
$^{5}$ARC Centre of Excellence for All Sky Astrophysics in 3 Dimensions (ASTRO 3D), Australia\\
$^{6}$National Astronomical Observatories, Chinese Academy of Sciences, A20 Datun Rd., Chaoyang District, Beijing, 100012, P.R. China\\
$^{7}$Department of Physics and Astronomy and PITT PACC, University of Pittsburgh, Pittsburgh, PA 15260, USA\\
$^{8}$Lawrence Berkeley National Laboratory, 1 Cyclotron Road, Berkeley, CA 94720, USA\\
$^{9}$Physics Dept., Boston University, 590 Commonwealth Avenue, Boston, MA 02215, USA\\
$^{10}$NSF NOIRLab, 950 N. Cherry Ave., Tucson, AZ 85719, USA\\
$^{11}$Instituto de F\'{\i}sica, Universidad Nacional Aut\'{o}noma de M\'{e}xico,  Cd. de M\'{e}xico  C.P. 04510,  M\'{e}xico\\
$^{12}$Departamento de F\'isica, Universidad de los Andes, Cra. 1 No. 18A-10, Edificio Ip, CP 111711, Bogot\'a, Colombia\\
$^{13}$Observatorio Astron\'omico, Universidad de los Andes, Cra. 1 No. 18A-10, Edificio H, CP 111711 Bogot\'a, Colombia\\
$^{14}$Institut d'Estudis Espacials de Catalunya (IEEC), 08034 Barcelona, Spain\\
$^{15}$Institute of Cosmology and Gravitation, University of Portsmouth, Dennis Sciama Building, Portsmouth, PO1 3FX, UK\\
$^{16}$Institute of Space Sciences, ICE-CSIC, Campus UAB, Carrer de Can Magrans s/n, 08913 Bellaterra, Barcelona, Spain\\
$^{17}$Department of Physics, Southern Methodist University, 3215 Daniel Avenue, Dallas, TX 75275, USA\\
$^{18}$Center for Cosmology and AstroParticle Physics, The Ohio State University, 191 West Woodruff Avenue, Columbus, OH 43210, USA\\
$^{19}$Department of Astronomy, The Ohio State University, 4055 McPherson Laboratory, 140 W 18th Avenue, Columbus, OH 43210, USA\\
$^{20}$The Ohio State University, Columbus, 43210 OH, USA\\
$^{21}$Instituci\'{o} Catalana de Recerca i Estudis Avan\c{c}ats, Passeig de Llu\'{\i}s Companys, 23, 08010 Barcelona, Spain\\
$^{22}$Institut de F\'{i}sica d’Altes Energies (IFAE), The Barcelona Institute of Science and Technology, Campus UAB, 08193 Bellaterra Barcelona, Spain\\
$^{23}$Department of Physics \& Astronomy, University  of Wyoming, 1000 E. University, Dept.~3905, Laramie, WY 82071, USA\\
$^{24}$Space Sciences Laboratory, University of California, Berkeley, 7 Gauss Way, Berkeley, CA  94720, USA\\
$^{25}$University of California, Berkeley, 110 Sproul Hall \#5800 Berkeley, CA 94720, USA\\
$^{26}$Department of Physics, Kansas State University, 116 Cardwell Hall, Manhattan, KS 66506, USA\\
$^{27}$Department of Physics and Astronomy, Sejong University, Seoul, 143-747, Korea\\
$^{28}$CIEMAT, Avenida Complutense 40, E-28040 Madrid, Spain\\
$^{29}$Department of Physics, University of Michigan, Ann Arbor, MI 48109, USA\\
$^{30}$University of Michigan, Ann Arbor, MI 48109, USA\\
$^{31}$Institute of Astronomy, University of Cambridge, Madingley Road, Cambridge CB3 0HA, UK\\
$^{32}$Instituto de Astrof\'{i}sica de Andaluc\'{i}a (CSIC), Glorieta de la Astronom\'{i}a, s/n, E-18008 Granada, Spain\\
}

\bsp	
\label{lastpage}
\end{document}